\documentclass[twoside,twocolumn,9pt]{article}
\usepackage{extsizes}
\usepackage[super,sort&compress,comma]{natbib}
\usepackage[version=4]{mhchem}
\usepackage[left=1.5cm, right=1.5cm, top=1.785cm, bottom=2.0cm]{geometry}
\usepackage{balance}
\usepackage{mathptmx}
\usepackage{sectsty}
\usepackage{graphicx}
\usepackage{lastpage}
\usepackage[format=plain,justification=justified,singlelinecheck=false,font={stretch=1.125,small,sf},labelfont=bf,labelsep=space]{caption}
\usepackage{float}
\usepackage{fancyhdr}
\usepackage{fnpos}
\usepackage[english]{babel}
\addto{\captionsenglish}{%
  
}
\usepackage{array}
\usepackage{droidsans}
\usepackage{charter}
\usepackage[T1]{fontenc}
\usepackage[usenames,dvipsnames,table]{xcolor}
\usepackage{setspace}
\usepackage[compact]{titlesec}
\usepackage{hyperref}

\usepackage{epstopdf}

\usepackage{amsmath}
\usepackage{amssymb}
\usepackage{rotating}
\usepackage{cleveref}
\usepackage{siunitx}
\graphicspath{{./}{rsc/}}

\definecolor{cream}{RGB}{222,217,201}

\newcommand*\Erkale{\textsc{Erkale}}
\newcommand*\HelFEM{\textsc{HelFEM}}
\newcommand*\OOO{\textsc{OpenOrbitalOptimizer}}

\newcommand*\citeref[1]{ref.~\citenum{#1}}

\begin{document}

\pagestyle{fancy}
\thispagestyle{plain}
\fancypagestyle{plain}{
\renewcommand{\headrulewidth}{0pt}
}

\makeFNbottom
\makeatletter
\renewcommand\LARGE{\@setfontsize\LARGE{15pt}{17}}
\renewcommand\Large{\@setfontsize\Large{12pt}{14}}
\renewcommand\large{\@setfontsize\large{10pt}{12}}
\renewcommand\footnotesize{\@setfontsize\footnotesize{7pt}{10}}
\makeatother

\renewcommand{\thefootnote}{\fnsymbol{footnote}}
\renewcommand\footnoterule{\vspace*{1pt}%
\color{cream}\hrule width 3.5in height 0.4pt \color{black}\vspace*{5pt}}
\setcounter{secnumdepth}{5}

\makeatletter
\renewcommand\@biblabel[1]{#1}
\renewcommand\@makefntext[1]%
{\noindent\makebox[0pt][r]{\@thefnmark\,}#1}
\makeatother
\renewcommand{\figurename}{\small{Fig.}~}
\sectionfont{\sffamily\Large}
\subsectionfont{\normalsize}
\subsubsectionfont{\bf}
\setstretch{1.125} 
\setlength{\skip\footins}{0.8cm}
\setlength{\footnotesep}{0.25cm}
\setlength{\jot}{10pt}
\titlespacing*{\section}{0pt}{4pt}{4pt}
\titlespacing*{\subsection}{0pt}{15pt}{1pt}

\fancyfoot{}
\fancyfoot[LO,RE]{\vspace{-7.1pt}\includegraphics[height=9pt]{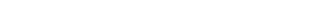}}
\fancyfoot[CO]{\vspace{-7.1pt}\hspace{11.9cm}\includegraphics{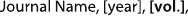}}
\fancyfoot[CE]{\vspace{-7.2pt}\hspace{-13.2cm}\includegraphics{head_foot/RF}}
\fancyfoot[RO]{\footnotesize{\sffamily{1--\pageref{LastPage} ~\textbar  \hspace{2pt}\thepage}}}
\fancyfoot[LE]{\footnotesize{\sffamily{\thepage~\textbar\hspace{4.65cm} 1--\pageref{LastPage}}}}
\fancyhead{}
\renewcommand{\headrulewidth}{0pt}
\renewcommand{\footrulewidth}{0pt}
\setlength{\arrayrulewidth}{1pt}
\setlength{\columnsep}{6.5mm}
\setlength\bibsep{1pt}

\makeatletter
\newlength{\figrulesep}
\setlength{\figrulesep}{0.5\textfloatsep}

\newcommand{\topfigrule}{\vspace*{-1pt}%
\noindent{\color{cream}\rule[-\figrulesep]{\columnwidth}{1.5pt}} }

\newcommand{\botfigrule}{\vspace*{-2pt}%
\noindent{\color{cream}\rule[\figrulesep]{\columnwidth}{1.5pt}} }

\newcommand{\dblfigrule}{\vspace*{-1pt}%
\noindent{\color{cream}\rule[-\figrulesep]{\textwidth}{1.5pt}} }

\makeatother

\twocolumn[
  \begin{@twocolumnfalse}
{\includegraphics[height=30pt]{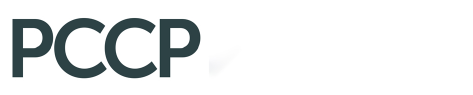}\hfill\raisebox{0pt}[0pt][0pt]{\includegraphics[height=55pt]{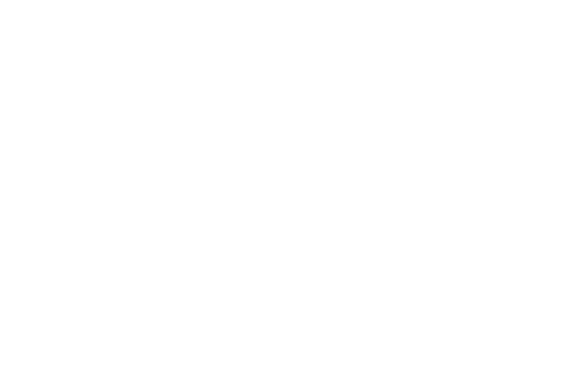}}\\[1ex]
\includegraphics[width=18.5cm]{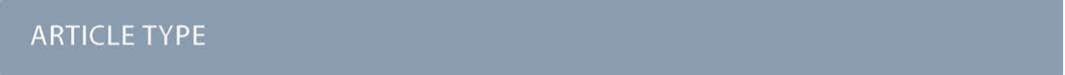}}\par
\vspace{1em}
\sffamily
\begin{tabular}{m{4.5cm} p{13.5cm} }

\includegraphics{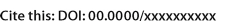} & \noindent\LARGE{\textbf{Real Quantum Chemistry With Complex Orbitals$^\dag$}} \\
\vspace{0.3cm} & \vspace{0.3cm} \\

 & \noindent\large{Hugo {\AA}str{\"o}m\textit{$^{a}$} and Susi Lehtola$^{\ast}$\textit{$^{a}$}} \\

\includegraphics{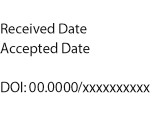} & \noindent\normalsize{We follow up our study of basis set truncation errors for atoms in magnetic fields [{\AA}str{\"o}m and Lehtola, \textit{J. Phys. Chem. A}, 2023, \textbf{127}, 10872]. Our previous study employed an approximate real-valued model. In this work, we implement a scheme to allow the use of complex basis functions and the true, complex Hamiltonian with linear molecules in a parallel magnetic field within the usual real-basis machinery of quantum chemistry. Our method performs additional unitary transformations before and after a conventional Fock build, thus allowing the reuse of existing software methods and algorithms. We apply our approach to calculations on low-lying configurations of the atoms $Z\leq18$ in static magnetic fields up to $0.6B_0$. The calculations employ the uncontracted aug-cc-pVTZ and the benchmarking quality AHGBSP3-9 Gaussian-type orbital basis sets. We compare total energies obtained with real and complex orbitals using these basis sets to fully numerical ones at the complete basis set limit. We see that the states of the real-valued Hamiltonian are superpositions of the true eigenstates that are correctly captured by the complex calculations. Our results show that the complex basis machinery is necessary for targeting states with the correct symmetry for the studied range of magnetic field strengths. The novel tool is key for future work where we aim to optimize basis sets for finite-field calculations.} \\

\end{tabular}

 \end{@twocolumnfalse} \vspace{0.6cm}

  ]

\renewcommand*\rmdefault{bch}\normalfont\upshape
\rmfamily
\section*{}
\vspace{-1cm}


\footnotetext{\textit{$^{a}$~University of Helsinki, Department of Chemistry, Faculty of Science, P.O. Box 55 (A.I. Virtanens plats 1), FI-00014 University of Helsinki, Finland. E-mail: susi.lehtola@alumni.helsinki.fi}}

\footnotetext{\dag~Electronic Supplementary Information (ESI) available: Tables of mean absolute energy differences for all electronic configurations of the H--Li and F--Na atoms in the real and complex aug-cc-pVTZ and AHGBSP3-9 basis sets, and figures with plots of energies of the electronic configurations of all atoms as functions of the magnetic field strength in the complex aug-cc-pVTZ and AHGBSP3-9 basis sets. See DOI: 10.1039/cXCP00000x/}



\section{Introduction}
\label{sec:introduction}

The electronic structure of atoms and molecules is strongly affected by magnetic fields of the order of one atomic unit ($1\ B_0\approx2.35\times10^5$ T).
The interesting intermediate regime $0.1\ B_0 \lesssim B \lesssim 10\ B_0$, where the electronic and magnetic interactions are of the same order of importance, can be found in white dwarf atmospheres,\cite{Greenstein1984_AJ_47, Greenstein1985_AJ_25, Hollands2023_MNRAS_3560} for example.
As these field strengths are outside the reach of experimental studies, electronic-structure calculations are needed for investigating chemistry in strong magnetic fields.
Such calculations have already led to the discovery of new chemistry, such as the discovery of the paramagnetic bonding mechanism of \ce{H2} and \ce{He2},\cite{Lange2012_S_327} and the structure of \ce{He} atom clusters in magnetic fields.\cite{Tellgren2012_PCCP_9492}

Calculating the electronic structure requires a suitable numerical discretization.
Atomic-orbital (AO) basis sets enjoy an overwhelming popularity in chemistry, with Gaussian-type orbital (GTO) basis sets being used first and foremost in the literature.
GTO basis sets are routinely employed also in finite-field calculations, and numerous studies at Hartree--Fock (HF),\cite{Jones1999_PRA_2875, Tellgren2008_JCP_154114,Tellgren2012_PCCP_9492, Sen2018_JCP_184112, Sun2018_JCTC_348, Culpitt2023_JCP_114115} density-functional theory (DFT),\cite{Furness2015_JCTC_4169,Reimann2017_JCTC_4089,WibowoTeale2024_PCCP_15156} coupled cluster (CC) theory,\cite{Hampe2017_JCP_154105, Hampe2019_JCTC_4036,Hampe2020_PCCP_23522, Stopkowicz2015_JCP_74110, Grazioli2025_JCTC_12634, Kitsaras2025_JCTC_10177, Blaschke2024_PCCP_28828, Kitsaras2024_JCP_094112} and configuration interaction\cite{Detmer1997_PRA_1825,Detmer1998_PRA_1767, Schmelcher1999_PRA_3424,Becken1999_JPBAMOP_1557, Becken2000_JPBAMOP_545,Becken2001_PRA_53412, AlHujaj2004_PRA_33411, AlHujaj2004_PRA_23411, Lange2012_S_327} (CI) levels of theory have been reported.
Yet, although an abundance of GTO basis sets is available for various purposes,\cite{Davidson1986_CR_681, Jensen2013_WIRCMS_273, Hill2013_IJQC_21} systematic basis sets have not yet been developed for calculations with strong magnetic fields to the best of our knowledge.

We initiated the work toward building such basis sets in a recent paper,\cite{Aastroem2023_JPCA_10872} wherein we studied the GTO basis set truncation errors (BSTEs) of low-lying configurations of the atoms $Z\leq18$ in static external magnetic fields up to $0.6B_0$.
We showed that standard GTO basis sets that are optimized at zero field are unable to capture the effects that the external field has on the electronic structure.
In contrast, the benchmarking-quality HGBS basis set family, and specifically the largest augmented polarized variant AHGBSP3-9 led to significant reduction of the errors.
Yet, some electronic configurations featured surprisingly large errors even in this extended basis set.
Furthermore, we saw configurations with unphysical negative BSTEs: the CBS energy is a lower bound on the GTO energy, so BSTEs should be non-negative.


Our previous work employed a real-valued approximation to the magnetic field Hamiltonian for compatibility with existing quantum chemistry machinery.
In this work, we introduce complex basis functions to our GTO code, backed by the realization that all integrals turn out real in the case of linear molecules in parallel fields.\cite{Lehtola2019_IJQC_25945, Lehtola2020_MP_1597989}
This realization allows us to build a complex-basis implementation on top of the usual real-basis machinery, with additional unitary transformations before and after the Fock build.
We will show in this work that the reason for the issues in our previous work is that our real-valued orbital approximation resulted in a mixing of configurations, and thereby a superposition of true eigenstates.
We also find that some of our previous CBS limit calculations had converged to saddle point solutions, and report improved CBS limit estimates in this work.

We will thus perform calculations on the previously studied electronic configurations of the atoms $Z\leq18$ in external fields up to $0.6B_0$ at the HF level.
We obtain complex-basis BSTEs and compare them to our real-basis results.
We assess the validity of the real-orbital approximation and resolve the unphysical results we found in \citeref{Aastroem2023_JPCA_10872}.

The layout of the paper is the following.
The theory is discussed in \cref{sec:theory}: we derive transformations between the real and complex basis in \cref{sec:c_trafo}, and outline our HF implementation in \cref{sec:implementation}.
We briefly discuss the computational details in \cref{sec:computational-details}, as they are essentially the same as in our previous work.\cite{Aastroem2023_JPCA_10872}
Then, we examine the results in \cref{sec:results}.
We see that the complex basis captures the correct symmetries, thus eliminating the negative BSTEs of excited states and further reducing other BSTEs.
We end with a summary and a discussion in \cref{sec:summary}.
Hartree atomic units are used unless otherwise specified.

\section{Theory \label{sec:theory}}

In electronic-structure calculations, the molecular orbitals (MOs), $\psi_i$, are traditionally expressed as a linear combination of basis functions $\{\chi_\alpha\}_{\alpha=1}^N$
\begin{equation}\label{eq:lcao}
\psi_i(\mathbf{r})=\sum_{\alpha=1}^N c_{\alpha i}\chi_\alpha(\mathbf{r}).
\end{equation}
The basis functions can take various forms but a popular choice is atomic orbitals (AOs), yielding the linear combination of atomic orbitals (LCAO) method.
The AOs, in turn, separate into a radial part and an angular part
\begin{equation}\label{eq:ao}
\chi_{nlm}(\mathbf{r})=R_{nl}(r)Y^m_{l}(\mathrm{\hat{r}}).
\end{equation}
The angular functions are the spherical harmonics, which can be either complex, denoted by $Y_l^m$, or real, denoted by $Y_{lm}$.
Both the real and complex spherical harmonics form complete basis sets on the sphere $S^2$, and they are connected by a unitary transform (see \cref{sec:c_trafo}).
Because the real functions avoid the need for complex algebra for general molecules, they are the common choice in quantum chemistry programs.
Complex basis functions can still be emulated within this choice by letting the molecular orbital coefficients ${\bf c}$ become complex.

The Hamiltonian for a linear molecule in a parallel magnetic field of strength $B$ on the $z$ axis can be written as\cite{Lehtola2020_MP_1597989}
\begin{equation}\label{eq:hamilt}
  \hat{H}=\hat{H}_0+\frac{1}{2}B\hat{L}_z+B\hat{S}_z+\frac{1}{8}B^2(x^2+y^2).
\end{equation}
The issue for finite-field calculations on linear molecules (which trivially include atoms and diatomic molecules) is that this operator depends on $L_z$, and only the complex spherical harmonic $Y_l^m$ is its eigenfunction (with eigenvalue $m$).
In contrast, in the real basis the $\hat{L}_z$ operator is purely imaginary and off-diagonal.

Our real-valued Hamiltonian approximation of \citeref{Aastroem2023_JPCA_10872}.
circumvented this issue by changing the Hamiltonian with the ansatz that $\hat{L}_z Y_{lm}=m Y_{lm}$.
The results of this approach differ in general from the physical one.
Our focus will be on the HF level of theory, as in our previous work in \citeref{Aastroem2023_JPCA_10872}.
The HF wave function is fully determined by the occupied orbitals.
Imposing symmetry constraints and the Aufbau principle, the occupied orbitals can be identified by their $m$ value.
The two approaches are then equivalent only for configurations with only $m=0$ orbitals occupied, or when the $m$ and $-m$ orbitals have equal occupation.

\subsection{Complex Transformations} \label{sec:c_trafo}

In this work, we describe the implementation of the physical Hamiltonian, but in a way that is compatible with real-basis machinery.
The connection between the real and complex spherical harmonics is
\begin{equation}\label{eq:sph_harm_trafo}
Y^m_l=
\begin{cases}
\displaystyle{\frac{1}{\sqrt{2}}\left(Y_{l,|m|}-iY_{l,-|m|}\right)},\quad m<0 \\
\displaystyle{Y_{l,m}},\quad m=0 \\
\displaystyle{\frac{(-1)^m}{\sqrt{2}}\left(Y_{l,|m|}+iY_{l,-|m|}\right)},\quad m>0.
\end{cases}
\end{equation}
We construct a linear transformation between the real and the complex AO basis from the relation in \cref{eq:sph_harm_trafo} as
\begin{equation} \label{eq:d_def}
\tilde{\chi}_\alpha(\mathbf{r})=\sum_\beta D_{\beta\alpha}\chi_\beta(\mathbf{r}),
\end{equation}
where ${\bf D}$ is a unitary matrix, ${\bf D} {\bf D}^\dagger = {\bf 1} = {\bf D}^\dagger {\bf D}$.

We will now be in a position to set up matrices in the real and the complex AO basis.
Matrices in the real AO basis such as the Fock matrix ${\bf F}$ are denoted without a tilde, while their counterparts in the complex basis are denoted with a tilde $\tilde{\bf F}$.
It is important to note that these matrices can be all be complex, a priori, since this leads to the most flexible ansatz for the orbital expansion in \cref{eq:lcao}.

Let us now assume that we have a MO in the complex AO basis as
\begin{equation}
  \psi_i({\bf r}) = \sum_{\alpha} \tilde{C}_{\alpha i} \tilde{\chi}_\alpha({\bf r}).
  \label{eq:psi-complex}
\end{equation}
We can find its expansion in the real AO basis as
\begin{equation} \begin{split}
\psi_i(\mathbf{r})&=\tilde{C}_{\alpha i}\tilde{\chi}_\alpha(\mathbf{r})=\tilde{C}_{\alpha i}D_{\beta\alpha}\chi_\beta(\mathbf{r}) \\
&=({\bf D}\tilde{\bf C})_{\beta i}\chi_\beta(\mathbf{r}) = C_{\beta i}\chi_\beta(\mathbf{r})
\end{split}
\end{equation}
from which we see that the MO coefficients transform as
\begin{equation} \label{eq:bas_mo}
{\bf C}={\bf D}\tilde{\bf C} \text{ and } \tilde{\bf C}={\bf D}^\dagger {\bf C}.
\end{equation}
It follows that the transformation property of the density matrix is
\begin{equation}
\begin{split}
\tilde{\bf P}&=\tilde{\bf C}{\bf n} \tilde{\bf C}^\dagger={\bf D}^\dagger {\bf C} {\bf n} ({\bf D}^\dagger {\bf C})^\dagger={\bf D}^\dagger {\bf C} {\bf n} {\bf C}^\dagger {\bf D} \\
&={\bf D}^\dagger {\bf PD},
\end{split}
\end{equation}
where ${\bf n}$ is the diagonal matrix of orbital occupations.

We will use the self-consistent field method to solve HF, which amounts to solving the Roothaan--Hall equation in the complex basis
\begin{equation}
\label{eq:complex-rh}
\tilde{\bf F}\tilde{\bf C}=\tilde{\bf S}\tilde{\bf C} {\bf E}.
\end{equation}
Since all integrals are real in this basis,\cite{Lehtola2019_IJQC_25945, Lehtola2020_MP_1597989} we especially know that $\tilde{\bf F}$ and $\tilde{\bf S}$ are real.
Now, we just need to derive the transformation of a matrix element $\langle i | \hat{O} | j \rangle$:
\begin{equation}
\label{eq:op-trans}
\begin{split}
 \langle i | \hat{O} | j \rangle &= \sum_{\alpha \beta} C^*_{\alpha i} O_{\alpha \beta} C_{\beta j} \\
 &= ({\bf D} \tilde{\bf C})^*_{\alpha i} O_{\alpha \beta} ({\bf D} \tilde{\bf C})_{\beta j} \\
 &= ({\bf D}^\dagger {\bf O} {\bf D})_{\alpha \beta} \tilde{\bf C}_{\alpha i} \tilde{\bf C}_{\beta j}.
 \end{split}
\end{equation}
\Cref{eq:op-trans} then lets us use the real-orbital AO machinery to build integrals for the complex AO basis, and thereby solve \cref{eq:complex-rh}: the complex-AO basis Fock matrix is obtained from the real-AO one by
\begin{equation} \label{eq:cbas_fock}
\tilde{\bf F}={\bf D}^\dagger {\bf F} {\bf D}.
\end{equation}

\section{Implementation} \label{sec:implementation}

\begin{figure}
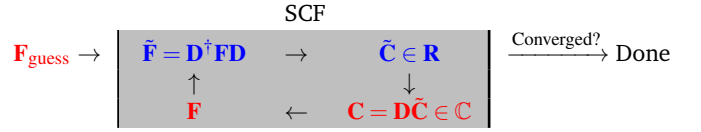

\centering
\begin{tabular}{r|ccc|l}
\multicolumn{1}{r}{} & \multicolumn{3}{c}{SCF} & \multicolumn{1}{l}{} \\
 {\color{red}{${\bf F}_\mathrm{guess}$}} $\rightarrow$ & \cellcolor{lightgray} \color{blue}{$\tilde{\bf F}={\bf D}^\dagger {\bf FD}$} & \cellcolor{lightgray} $\rightarrow$ & \cellcolor{lightgray} \color{blue}{$\tilde{\bf C}\in\mathbb{\bf R}$} & $\xrightarrow{\mathrm{Converged?}}$ Done \\
  & \cellcolor{lightgray} $\uparrow$ & \cellcolor{lightgray} & \cellcolor{lightgray} $\downarrow$ & \\
  & \cellcolor{lightgray} \color{red}{${\bf F}$} & \cellcolor{lightgray} $\leftarrow$ & \cellcolor{lightgray} \color{red}{${\bf C}={\bf D}\tilde{\bf C}\in\mathbb{C}$} & \\
\end{tabular}
\caption{Outline of the algorithm.
Steps carried out in the real basis and complex basis are written in {\color{red}{red}} and {\color{blue}{blue}}, respectively.}
\label{fig:alg}
\end{figure}

Following established quantum chemistry methodology, all molecular integrals are evaluated in the real-valued AO basis, while only the $\hat{L}_z$ interaction term is evaluated directly in the complex AO basis.
The procedure is visualized in \cref{fig:alg}.
For HF calculations, we need two ingredients: the electron-electron Coulomb repulsion matrix
\begin{equation}
J_{\mu\nu}=\sum_{\kappa\lambda}P_{\kappa\lambda}(\mu\nu|\kappa\lambda)
\end{equation}
as well as the exchange matrix
\begin{equation}
K_{\mu\nu}=\sum_{\kappa\lambda}P_{\kappa\lambda}(\kappa\mu|\lambda\nu).
\end{equation}

For real basis functions it holds that
\begin{equation}
(\mu\nu|\kappa\lambda)=(\mu\nu|\lambda\kappa)
\end{equation}
is symmetric in $(\kappa,\lambda)$.
Since ${\bf P}$ is Hermitian,  $\textrm{Re} ({\bf P})$ is also symmetric, while $\mathrm{Im}({\bf P})$ is antisymmetric.
It follows that the contraction of $\mathrm{Im}({\bf P})$ with $(\mu\nu|\kappa\lambda)$ vanishes and only $\mathrm{Re}({\bf P})$ contributes to ${\bf J}$.
In contrast, the exchange matrix ${\bf K}$ gets a contribution from the imaginary part of the density matrix.
However, in the complex basis $\tilde{\bf K}$ still turns out to be real, since $\tilde{\bf P}$ is real, and the two-electron integrals are real in the complex basis as well.\cite{Lehtola2019_IJQC_25945}

The Coulomb build thus can be done with a traditional call, while the complex exchange build can either employ separate builds for the real and imaginary parts arising from the real and imaginary parts of the density matrix, or a complex-orbital extension as already available in the \Erkale{} program\cite{Lehtola2019_IJQC_25945} where the present method has been implemented.

\section{Computational Details}
\label{sec:computational-details}

We adopt the methods of \citeref{Aastroem2023_JPCA_10872}.
The GTO calculations are performed with \Erkale{}\cite{Lehtola2012_JCC_1572} with the correlation consistent aug-cc-pVTZ basis set\cite{Dunning1989_JCP_1007, Kendall1992_JCP_6796, Woon1993_JCP_1358,Peterson1994_JCP_7410} in its fully uncontracted form, as well as the hydrogenic Gaussian basis set AHGBSP3-9.\cite{Lehtola2020_JCP_134108}
The self-consistent field solution in the GTO calculations of this work is handled by \OOO.\cite{Lehtola2025_JPCA_5651, Lehtola2025__}
The fully numerical calculations are performed with the open source \HelFEM{} software,\cite{Lehtola2019_IJQC_25945, Lehtola2019_IJQC_25944, Lehtola2020_MP_1597989, HelFEM} and they are converged to the complete basis set (CBS) limit.
All software is open source\cite{Lehtola2022_WIRCMS_1610} and freely available on GitHub.

All calculations are performed at the unrestricted HF (UHF) level of theory.
We take the low-lying configurations of the atoms $Z\leq18$ from \citeref{Aastroem2023_JPCA_10872} and perform the GTO calculations in external magnetic fields $B\in\{0.00,0.02,\dots,0.60\}B_0$ with complex orbitals.
The corresponding field strengths in the fully numerical calculations are $B\in\{0.00,0.10,\dots,0.60\}B_0$.
The reader is referred to our previous work for a detailed discussion on the configuration search and the parameters of the fully numerical calculations.

We make some further remarks on the initial guess.
In \citeref{Aastroem2023_JPCA_10872} we used a superposition of atomic potentials\cite{Lehtola2019_JCTC_1593, Lehtola2020_JCP_144105} (SAP).
We found in this work that some of the earlier calculations had silently converged to saddle point solutions.
We therefore performed further calculations with the core guess in this work.
The core guess was found to also lead to some saddle point solutions, but different ones than with SAP.
In some difficult cases, neither of these guesses led to a converged solution.
In these cases we used the converged Fock matrix from a stronger or weaker magnetic field as initial guess.
This was also done in similar cases in the previous work.
The final results are obtained by choosing the calculation with the lowest energy.
This leads to smooth behavior of the studied configurations' energies.

Note that polyatomic GTO calculations typically employ the gauge-including atomic orbital (GIAO) also known as London orbital approach.\cite{London1937_JPlR_397, Hameka1958_MP_203, Ditchfield1974_MP_789, Wolinski1990_JACS_8251}
The GIAO phase factor $\exp (i {\bf B} \times {\bf R}_{\mu} \cdot {\bf r})$ vanishes when ${\bf B} \parallel {\bf R}_\mu$, which is the case for atoms and linear molecules.
The results of this work are then equivalent to the use of GIAO, alike our previous study on atoms\cite{Aastroem2023_JPCA_10872} and diatomic molecules.\cite{Lehtola2020_MP_1597989}

\section{Results}
\label{sec:results}

To keep a compact notation, we follow our previous work and denote the studied electronic configurations as
\begin{equation} \label{eq:label}
\prod_{m\in\{\sigma,\pi,\delta,\phi\}} m_{+/-}^{n_\alpha,n_\beta},
\end{equation}
where $\sigma,\pi,\delta,\phi$ indicate $m=0,1,2,3$, respectively, $+/-$ indicates the sign of $m$, and $n_\alpha$ and $n_\beta$ are the number of $\alpha$ and $\beta$ electrons occupying orbitals with this value of $m$.
For example, the zero-field UHF ground state of the F atom with orbital occupations $1\sigma^{2}2\sigma^{2}3\sigma^{1}1\pi_+^{2}1\pi_-^{2}$ is now denoted as $\sigma^{3,2}\pi_+^{1,1}\pi_-^{1,1}$.
Note that the $\alpha$ and $\beta$ spatial orbitals may differ in UHF.

The BSTE at field strength $B$ is defined as
\begin{equation} \label{eq:bste}
\Delta E^\mathrm{GTO}(B)=E^\mathrm{GTO}(B)-E^\mathrm{CBS}(B),
\end{equation}
where $E^\mathrm{CBS}$ is the fully numerical energy converged to the CBS limit.
We average the absolute BSTEs and define the mean absolute energy difference (MAED) as
\begin{equation}
\Delta E^\text{GTO} =\frac 1 N \sum_{i=1}^N |\Delta E^\text{GTO} (B_i)|, \label{eq:maed}
\end{equation}
where $N=7$ and $B_i\in\{0.00,0.10,\dots,0.60\}B_0$.
The MAEDs form the basis of our analysis.

The analysis is further aided with plots of the energy as functions of the magnetic field strength for each electronic configuration.
These reveal the rich chemistry emerging from the magnetic field, as discussed in detail in the previous work.\cite{Aastroem2023_JPCA_10872}
The behavior is smooth and qualitatively the same as the real-basis behavior.
The only visible difference is that the complex energy curves appear to agree better with the CBS energies for select electronic configurations.
The plots are found in the ESI.\dag

We proceed with an analysis of the differences between the real and complex MAEDs.
In the following, electronic configuration labels highlighted in magenta represent states that are equivalent in the real and complex basis.
We see that their real and complex MAEDs are identical, as expected.
Also, table entries in a light red color belong to states that feature negative BSTEs at one or more magnetic field strengths.
All negative BSTEs observed with the real basis are eliminated with the complex basis.

\paragraph{H--Li.~~}
For the H, He, and Li atoms all real- and complex-basis MAEDs are identical and the tables containing them can be found in the ESI.\dag
The only exception is the $\sigma^{2,1}$ state of Li.
We would have expected this state to also be identical, since only $\sigma$-orbitals are occupied; the disagreement arises because the core guess results in convergence to a lower energy than the SAP guess from the previous work at $B=0.60B_0$.

\paragraph{Be--O.~~}
The MAEDs for the Be--O atoms are included in \cref{tab:Be-mean-differ,tab:B-mean-differ,tab:C-mean-differ,tab:N-mean-differ,tab:O-mean-differ}.
With the real AHGBSP3-9 basis we see a clear pattern, where the MAEDs of one or two electronic configurations are unexpectedly large.
These states are $\sigma^{1,1}\pi_-^{1,0}\delta_-^{1,0}$ of the Be atom, $\sigma^{2,1}\pi_-^{1,0}\delta_-^{1,0}$ of B, $\sigma^{3,1}\pi_-^{1,0}\delta_-^{1,0}$ of the C atom, $\sigma^{3,2}\pi_+^{1,0}$ and $\sigma^{3,2}\pi_-^{1,0}$ of N, and $\sigma^{3,3}\pi_-^{1,1}$ of the O atom.
With the complex orbitals these MAEDs are reduced by one to four orders of magnitude.
In the aug-cc-pVTZ basis we observe a similar but less pronounced trend than that of the AHGBSP3-9 basis.
The other states show either no change, or a small reduction in the MAEDs when switching from the real to the complex basis, as can be seen in the tables.
We can also see that configurations occupying orbitals with $m=3$ are difficult to describe, even with the complex basis.
Still, the MAEDs in the AHGBSP3-9 basis are one to two orders of magnitude smaller than in the aug-cc-pVTZ basis.
Therefore, we can simply attribute these to the basis set truncation.

\begin{table*}
\centering
\small
\begin{tabular}{c|S[table-format=4.3]S[table-format=4.3]S[table-format=4.3]|S[table-format=4.3]S[table-format=4.3]S[table-format=4.3]}
\hline
 & \multicolumn{3}{|c}{aug-cc-pVTZ} & \multicolumn{3}{|c}{AHGBSP3-9}\\
\hline \hline
state & {Real} & {Complex} & {Change} & {Real} & {Complex} & {Change}\\ 
\hline \hline
\color{magenta}{$\sigma^{2,2}$} & 0.902 & 0.902 & 0.000 & 0.034 & 0.034 & 0.000\\ 
\color{black}{$\sigma^{2,1}\pi_-^{1,0}$} & 1.150 & 1.150 & 0.000 & 0.057 & 0.057 & 0.000\\ 
\color{black}{$\sigma^{2,1}\pi_+^{1,0}$} & 1.150 & 1.150 & 0.000 & 0.057 & 0.057 & 0.000\\ 
\color{magenta}{$\sigma^{3,1}$} & 1.651 & 1.651 & 0.000 & 0.221 & 0.221 & 0.000\\ 
\color{black}{$\sigma^{2,1}\delta_-^{1,0}$} & 21.312 & 21.312 & 0.000 & 0.782 & 0.782 & 0.000\\ 
\color{black}{$\sigma^{1,1}\pi_-^{1,0}\delta_-^{1,0}$} & 36.338 & 18.182 & -18.156 & 16.129 & 0.622 & -15.506\\ 
\hline
\end{tabular}
\caption{MAEDs between GTO and FEM energies in m$E_h$ for electronic configurations (following our shorthand notation) of the Be atom in the fully uncontracted basis sets with real and complex orbitals. Configurations highlighted in magenta are equal in the real and complex basis and entries in light red belong to configurations that feature negative BSTEs.}
\label{tab:Be-mean-differ}
\end{table*}

\begin{table*}
\centering
\small
\begin{tabular}{c|S[table-format=4.3]S[table-format=4.3]S[table-format=4.3]|S[table-format=4.3]S[table-format=4.3]S[table-format=4.3]}
\hline
 & \multicolumn{3}{|c}{aug-cc-pVTZ} & \multicolumn{3}{|c}{AHGBSP3-9}\\
\hline \hline
state & {Real} & {Complex} & {Change} & {Real} & {Complex} & {Change}\\ 
\hline \hline
\color{black}{$\sigma^{2,2}\pi_+^{1,0}$} & 1.523 & 1.523 & 0.000 & 0.015 & 0.015 & 0.000\\ 
\color{black}{$\sigma^{2,2}\pi_-^{1,0}$} & 1.523 & 1.523 & 0.000 & 0.015 & 0.015 & 0.000\\ 
\color{magenta}{$\sigma^{2,1}\pi_+^{1,0}\pi_-^{1,0}$} & 1.780 & 1.780 & 0.000 & 0.022 & 0.022 & 0.000\\ 
\color{black}{$\sigma^{3,1}\pi_-^{1,0}$} & 1.855 & 1.855 & 0.000 & 0.055 & 0.055 & 0.000\\ 
\color{black}{$\sigma^{2,1}\pi_-^{1,0}\delta_-^{1,0}$} & 34.409 & 19.853 & -14.556 & 12.586 & 0.491 & -12.096\\ 
\color{black}{$\sigma^{2,1}\pi_-^{2,0}$} & 32.947 & 28.897 & -4.050 & 6.849 & 3.567 & -3.282\\ 
\hline
\end{tabular}
\caption{MAEDs between GTO and FEM energies in m$E_h$ for electronic configurations (following our shorthand notation) of the B atom in the fully uncontracted basis sets with real and complex orbitals. Configurations highlighted in magenta are equal in the real and complex basis and entries in light red belong to configurations that feature negative BSTEs.}
\label{tab:B-mean-differ}
\end{table*}

\begin{table*}
\centering
\small
\begin{tabular}{c|S[table-format=4.3]S[table-format=4.3]S[table-format=4.3]|S[table-format=4.3]S[table-format=4.3]S[table-format=4.3]}
\hline
 & \multicolumn{3}{|c}{aug-cc-pVTZ} & \multicolumn{3}{|c}{AHGBSP3-9}\\
\hline \hline
state & {Real} & {Complex} & {Change} & {Real} & {Complex} & {Change}\\ 
\hline \hline
\color{magenta}{$\sigma^{2,2}\pi_+^{1,0}\pi_-^{1,0}$} & 2.761 & 2.761 & 0.000 & 0.006 & 0.006 & 0.000\\ 
\color{black}{$\sigma^{3,2}\pi_+^{1,0}$} & 2.714 & 2.714 & 0.000 & 0.016 & 0.016 & 0.000\\ 
\color{black}{$\sigma^{3,2}\pi_-^{1,0}$} & 2.714 & 2.714 & 0.000 & 0.016 & 0.016 & 0.000\\ 
\color{magenta}{$\sigma^{3,1}\pi_+^{1,0}\pi_-^{1,0}$} & 3.288 & 3.288 & 0.000 & 0.015 & 0.015 & 0.000\\ 
\color{black}{$\sigma^{3,1}\pi_-^{1,0}\delta_-^{1,0}$} & 51.333 & 39.416 & -11.917 & 9.391 & 0.457 & -8.934\\ 
\color{black}{$\sigma^{3,1}\pi_-^{1,0}\phi_-^{1,0}$} & 452.152 & 447.025 & -5.127 & 12.204 & 11.387 & -0.817\\ 
\hline
\end{tabular}
\caption{MAEDs between GTO and FEM energies in m$E_h$ for electronic configurations (following our shorthand notation) of the C atom in the fully uncontracted basis sets with real and complex orbitals. Configurations highlighted in magenta are equal in the real and complex basis and entries in light red belong to configurations that feature negative BSTEs.}
\label{tab:C-mean-differ}
\end{table*}

\begin{table*}
\centering
\small
\begin{tabular}{c|S[table-format=4.3]S[table-format=4.3]S[table-format=4.3]|S[table-format=4.3]S[table-format=4.3]S[table-format=4.3]}
\hline
 & \multicolumn{3}{|c}{aug-cc-pVTZ} & \multicolumn{3}{|c}{AHGBSP3-9}\\
\hline \hline
state & {Real} & {Complex} & {Change} & {Real} & {Complex} & {Change}\\ 
\hline \hline
\color{magenta}{$\sigma^{3,2}\pi_+^{1,0}\pi_-^{1,0}$} & 4.147 & 4.147 & 0.000 & 0.005 & 0.005 & 0.000\\ 
\color{magenta}{$\sigma^{2,3}\pi_+^{1,0}\pi_-^{1,0}$} & 4.438 & 4.438 & 0.000 & 0.008 & 0.008 & 0.000\\ 
\color{black}{$\sigma^{3,2}\pi_+^{1,1}$} & 39.321 & 4.203 & -35.118 & 35.137 & 0.007 & -35.130\\ 
\color{black}{$\sigma^{3,2}\pi_-^{1,1}$} & 39.321 & 4.203 & -35.118 & 35.137 & 0.007 & -35.130\\ 
\color{black}{$\sigma^{3,3}\pi_-^{1,0}$} & 4.129 & 4.129 & 0.000 & 0.011 & 0.011 & 0.000\\ 
\color{black}{$\sigma^{3,1}\pi_+^{1,0}\pi_-^{1,0}\delta_-^{1,0}$} & 81.788 & 81.404 & -0.384 & 0.802 & 0.627 & -0.174\\ 
\color{black}{$\sigma^{3,1}\pi_+^{1,0}\pi_-^{1,0}\phi_-^{1,0}$} & 760.601 & 760.464 & -0.137 & 12.395 & 12.389 & -0.006\\ 
\hline
\end{tabular}
\caption{MAEDs between GTO and FEM energies in m$E_h$ for electronic configurations (following our shorthand notation) of the N atom in the fully uncontracted basis sets with real and complex orbitals. Configurations highlighted in magenta are equal in the real and complex basis and entries in light red belong to configurations that feature negative BSTEs.}
\label{tab:N-mean-differ}
\end{table*}

\begin{table*}
\centering
\small
\begin{tabular}{c|S[table-format=4.3]S[table-format=4.3]S[table-format=4.3]|S[table-format=4.3]S[table-format=4.3]S[table-format=4.3]}
\hline
 & \multicolumn{3}{|c}{aug-cc-pVTZ} & \multicolumn{3}{|c}{AHGBSP3-9}\\
\hline \hline
state & {Real} & {Complex} & {Change} & {Real} & {Complex} & {Change}\\ 
\hline \hline
\color{magenta}{$\sigma^{3,3}\pi_+^{1,0}\pi_-^{1,0}$} & 6.769 & 6.769 & 0.000 & 0.004 & 0.004 & 0.000\\ 
\color{black}{$\sigma^{3,2}\pi_+^{1,1}\pi_-^{1,0}$} & \cellcolor{pink}6.426 & 6.767 & 0.341 & \cellcolor{pink}0.340 & 0.003 & -0.337\\ 
\color{black}{$\sigma^{3,2}\pi_+^{1,0}\pi_-^{1,1}$} & \cellcolor{pink}6.426 & 6.767 & 0.341 & \cellcolor{pink}0.340 & 0.003 & -0.337\\ 
\color{black}{$\sigma^{3,3}\pi_-^{1,1}$} & 47.676 & 6.580 & -41.096 & 41.149 & 0.004 & -41.145\\ 
\color{black}{$\sigma^{3,2}\pi_+^{1,0}\pi_-^{1,0}\delta_-^{1,0}$} & 169.531 & 169.101 & -0.431 & 0.901 & 0.795 & -0.106\\ 
\color{black}{$\sigma^{3,2}\pi_+^{1,0}\pi_-^{2,0}$} & \cellcolor{pink}46.803 & 46.878 & 0.075 & \cellcolor{pink}4.010 & 4.057 & 0.047\\ 
\color{black}{$\sigma^{3,2}\pi_+^{1,0}\pi_-^{1,0}\phi_-^{1,0}$} & 1219.923 & 1219.772 & -0.151 & 12.693 & 12.690 & -0.003\\ 
\hline
\end{tabular}
\caption{MAEDs between GTO and FEM energies in m$E_h$ for electronic configurations (following our shorthand notation) of the O atom in the fully uncontracted basis sets with real and complex orbitals. Configurations highlighted in magenta are equal in the real and complex basis and entries in light red belong to configurations that feature negative BSTEs.}
\label{tab:O-mean-differ}
\end{table*}

\paragraph{F--Na.~~}
The tables for the F--Na atoms can be found in the ESI.\dag
All their MAEDs are qualitatively similar with the real and the complex orbitals in both the aug-cc-pVTZ and AHGBSP3-9 basis sets.
Slight reductions in the MAEDs can however be observed for the $\sigma^{3,2}\pi_+^{1,0}\pi_-^{2,1}$ configuration of F and the $\sigma^{3,3}\pi_+^{1,0}\pi_-^{2,1}$ configuration of Ne when switching to the complex basis.

\paragraph{Mg--Al.~~}
The results for the Mg and Al atoms are found in \cref{tab:Mg-mean-differ,tab:Al-mean-differ} and they follow the trend of the Be--O atoms.
In the real AHGBSP3-9 basis the $\sigma^{3,3}\pi_+^{1,1}\pi_-^{2,1}\delta_-^{1,0}$ configuration of the Mg atom, and the $\sigma^{4,3}\pi_+^{1,1}\pi_-^{2,1}\delta_-^{1,0}$ configuration of Al feature noticeable MAEDs compared to the other configurations.
These are reduced by two to three orders of magnitude with the complex basis.
In the aug-cc-pVTZ basis the corresponding reduction is about a factor of two.
The other states exhibit qualitatively similar, albeit slightly smaller MAEDs when going from the real to the complex basis.

\begin{table*}
\centering
\small
\begin{tabular}{c|S[table-format=4.3]S[table-format=4.3]S[table-format=4.3]|S[table-format=4.3]S[table-format=4.3]S[table-format=4.3]}
\hline
 & \multicolumn{3}{|c}{aug-cc-pVTZ} & \multicolumn{3}{|c}{AHGBSP3-9}\\
\hline \hline
state & {Real} & {Complex} & {Change} & {Real} & {Complex} & {Change}\\ 
\hline \hline
\color{magenta}{$\sigma^{4,4}\pi_+^{1,1}\pi_-^{1,1}$} & 3.261 & 3.261 & 0.000 & 0.186 & 0.186 & 0.000\\ 
\color{magenta}{$\sigma^{5,3}\pi_+^{1,1}\pi_-^{1,1}$} & 4.436 & 4.436 & 0.000 & 1.283 & 1.283 & 0.000\\ 
\color{black}{$\sigma^{4,3}\pi_+^{2,1}\pi_-^{1,1}$} & \cellcolor{pink}3.202 & 3.239 & 0.037 & \cellcolor{pink}0.430 & 0.456 & 0.025\\ 
\color{black}{$\sigma^{4,3}\pi_+^{1,1}\pi_-^{2,1}$} & \cellcolor{pink}3.202 & 3.239 & 0.037 & \cellcolor{pink}0.430 & 0.456 & 0.025\\ 
\color{black}{$\sigma^{4,3}\pi_+^{1,1}\pi_-^{1,1}\delta_-^{1,0}$} & 16.205 & 16.190 & -0.015 & 0.547 & 0.528 & -0.019\\ 
\color{black}{$\sigma^{3,3}\pi_+^{1,1}\pi_-^{2,1}\delta_-^{1,0}$} & 37.781 & 14.037 & -23.744 & 23.214 & 0.686 & -22.528\\ 
\hline
\end{tabular}
\caption{MAEDs between GTO and FEM energies in m$E_h$ for electronic configurations (following our shorthand notation) of the Mg atom in the fully uncontracted basis sets with real and complex orbitals. Configurations highlighted in magenta are equal in the real and complex basis and entries in light red belong to configurations that feature negative BSTEs.}
\label{tab:Mg-mean-differ}
\end{table*}

\begin{table*}
\centering
\small
\begin{tabular}{c|S[table-format=4.3]S[table-format=4.3]S[table-format=4.3]|S[table-format=4.3]S[table-format=4.3]S[table-format=4.3]}
\hline
 & \multicolumn{3}{|c}{aug-cc-pVTZ} & \multicolumn{3}{|c}{AHGBSP3-9}\\
\hline \hline
state & {Real} & {Complex} & {Change} & {Real} & {Complex} & {Change}\\ 
\hline \hline
\color{magenta}{$\sigma^{5,4}\pi_+^{1,1}\pi_-^{1,1}$} & 4.016 & 4.016 & 0.000 & 0.576 & 0.576 & 0.000\\ 
\color{black}{$\sigma^{4,4}\pi_+^{2,1}\pi_-^{1,1}$} & \cellcolor{pink}3.474 & 3.510 & 0.036 & \cellcolor{pink}0.139 & 0.158 & 0.018\\ 
\color{black}{$\sigma^{4,4}\pi_+^{1,1}\pi_-^{2,1}$} & \cellcolor{pink}3.474 & 3.510 & 0.036 & \cellcolor{pink}0.139 & 0.158 & 0.018\\ 
\color{black}{$\sigma^{5,3}\pi_+^{1,1}\pi_-^{2,1}$} & \cellcolor{pink}4.925 & 4.965 & 0.040 & \cellcolor{pink}0.464 & 0.484 & 0.020\\ 
\color{magenta}{$\sigma^{4,3}\pi_+^{2,1}\pi_-^{2,1}$} & 4.698 & 4.698 & 0.000 & 0.235 & 0.235 & 0.000\\ 
\color{black}{$\sigma^{4,3}\pi_+^{1,1}\pi_-^{2,1}\delta_-^{1,0}$} & 41.080 & 16.519 & -24.560 & 24.224 & 0.252 & -23.972\\ 
\color{black}{$\sigma^{4,3}\pi_+^{1,1}\pi_-^{3,1}$} & 25.320 & 20.988 & -4.332 & 6.287 & 1.503 & -4.784\\ 
\hline
\end{tabular}
\caption{MAEDs between GTO and FEM energies in m$E_h$ for electronic configurations (following our shorthand notation) of the Al atom in the fully uncontracted basis sets with real and complex orbitals. Configurations highlighted in magenta are equal in the real and complex basis and entries in light red belong to configurations that feature negative BSTEs.}
\label{tab:Al-mean-differ}
\end{table*}

\paragraph{Si--Ar.~~}
The MAEDs for the Si to Ar atoms are found in \cref{tab:Si-mean-differ,tab:P-mean-differ,tab:S-mean-differ,tab:Cl-mean-differ,tab:Ar-mean-differ}.
We observe that the majority of the states exhibit significant MAEDs when employing real orbitals, even with the AHGBSP3-9 basis.
However, going to the complex basis reduces these by one to two orders of magnitude.
The differences with the aug-cc-pVTZ basis are again more modest but still clear, as can be seen in the tables.

\begin{table*}
\centering
\small
\begin{tabular}{c|S[table-format=4.3]S[table-format=4.3]S[table-format=4.3]|S[table-format=4.3]S[table-format=4.3]S[table-format=4.3]}
\hline
 & \multicolumn{3}{|c}{aug-cc-pVTZ} & \multicolumn{3}{|c}{AHGBSP3-9}\\
\hline \hline
state & {Real} & {Complex} & {Change} & {Real} & {Complex} & {Change}\\ 
\hline \hline
\color{magenta}{$\sigma^{4,4}\pi_+^{2,1}\pi_-^{2,1}$} & 4.342 & 4.342 & 0.000 & 0.085 & 0.085 & 0.000\\ 
\color{black}{$\sigma^{5,4}\pi_+^{2,1}\pi_-^{1,1}$} & \cellcolor{pink}4.019 & 4.058 & 0.039 & \cellcolor{pink}0.203 & 0.218 & 0.015\\ 
\color{black}{$\sigma^{5,4}\pi_+^{1,1}\pi_-^{2,1}$} & \cellcolor{pink}4.019 & 4.058 & 0.039 & \cellcolor{pink}0.203 & 0.218 & 0.015\\ 
\color{magenta}{$\sigma^{5,3}\pi_+^{2,1}\pi_-^{2,1}$} & 4.743 & 4.743 & 0.000 & 0.186 & 0.186 & 0.000\\ 
\color{black}{$\sigma^{5,3}\pi_+^{1,1}\pi_-^{2,1}\delta_-^{1,0}$} & 43.846 & 16.880 & -26.965 & 26.202 & 0.212 & -25.990\\ 
\color{black}{$\sigma^{5,3}\pi_+^{1,1}\pi_-^{3,1}$} & 23.209 & 18.135 & -5.074 & 6.468 & 0.579 & -5.889\\ 
\color{black}{$\sigma^{4,3}\pi_+^{1,1}\pi_-^{3,1}\delta_-^{1,0}$} & 74.118 & 34.466 & -39.652 & 40.056 & 0.768 & -39.287\\ 
\color{black}{$\sigma^{4,3}\pi_+^{1,1}\pi_-^{2,1}\delta_-^{1,0}\phi_-^{1,0}$} & 157.267 & 93.672 & -63.595 & 61.168 & 8.129 & -53.039\\ 
\hline
\end{tabular}
\caption{MAEDs between GTO and FEM energies in m$E_h$ for electronic configurations (following our shorthand notation) of the Si atom in the fully uncontracted basis sets with real and complex orbitals. Configurations highlighted in magenta are equal in the real and complex basis and entries in light red belong to configurations that feature negative BSTEs.}
\label{tab:Si-mean-differ}
\end{table*}

\begin{table*}
\centering
\small
\begin{tabular}{c|S[table-format=4.3]S[table-format=4.3]S[table-format=4.3]|S[table-format=4.3]S[table-format=4.3]S[table-format=4.3]}
\hline
 & \multicolumn{3}{|c}{aug-cc-pVTZ} & \multicolumn{3}{|c}{AHGBSP3-9}\\
\hline \hline
state & {Real} & {Complex} & {Change} & {Real} & {Complex} & {Change}\\ 
\hline \hline
\color{magenta}{$\sigma^{5,4}\pi_+^{2,1}\pi_-^{2,1}$} & 3.930 & 3.930 & 0.000 & 0.085 & 0.085 & 0.000\\ 
\color{magenta}{$\sigma^{4,5}\pi_+^{2,1}\pi_-^{2,1}$} & 4.585 & 4.585 & 0.000 & 0.126 & 0.126 & 0.000\\ 
\color{black}{$\sigma^{5,4}\pi_+^{1,2}\pi_-^{2,1}$} & \cellcolor{pink}20.651 & 4.177 & -16.474 & \cellcolor{pink}24.685 & 0.115 & -24.570\\ 
\color{black}{$\sigma^{5,4}\pi_+^{1,1}\pi_-^{2,2}$} & 28.453 & 4.177 & -24.277 & 24.353 & 0.115 & -24.238\\ 
\color{black}{$\sigma^{5,4}\pi_+^{1,1}\pi_-^{2,1}\delta_-^{1,0}$} & 49.091 & 21.267 & -27.823 & 25.662 & 0.120 & -25.543\\ 
\color{black}{$\sigma^{5,3}\pi_+^{2,1}\pi_-^{2,1}\delta_-^{1,0}$} & 21.605 & 21.338 & -0.267 & 0.309 & 0.110 & -0.199\\ 
\color{black}{$\sigma^{5,3}\pi_+^{1,1}\pi_-^{3,1}\delta_-^{1,0}$} & 82.647 & 39.149 & -43.498 & 43.272 & 0.304 & -42.968\\ 
\color{black}{$\sigma^{5,3}\pi_+^{1,1}\pi_-^{2,1}\delta_-^{1,0}\phi_-^{1,0}$} & 213.458 & 144.298 & -69.161 & 60.514 & 6.744 & -53.769\\ 
\color{black}{$\sigma^{6,3}\pi_+^{1,1}\pi_-^{2,1}\delta_-^{1,0}$} & 73.720 & 40.648 & -33.072 & 35.840 & 1.977 & -33.863\\ 
\hline
\end{tabular}
\caption{MAEDs between GTO and FEM energies in m$E_h$ for electronic configurations (following our shorthand notation) of the P atom in the fully uncontracted basis sets with real and complex orbitals. Configurations highlighted in magenta are equal in the real and complex basis and entries in light red belong to configurations that feature negative BSTEs.}
\label{tab:P-mean-differ}
\end{table*}

\begin{table*}
\centering
\small
\begin{tabular}{c|S[table-format=4.3]S[table-format=4.3]S[table-format=4.3]|S[table-format=4.3]S[table-format=4.3]S[table-format=4.3]}
\hline
 & \multicolumn{3}{|c}{aug-cc-pVTZ} & \multicolumn{3}{|c}{AHGBSP3-9}\\
\hline \hline
state & {Real} & {Complex} & {Change} & {Real} & {Complex} & {Change}\\ 
\hline \hline
\color{magenta}{$\sigma^{5,5}\pi_+^{2,1}\pi_-^{2,1}$} & 4.813 & 4.813 & 0.000 & 0.065 & 0.065 & 0.000\\ 
\color{black}{$\sigma^{5,4}\pi_+^{2,1}\pi_-^{2,2}$} & \cellcolor{pink}4.399 & 4.641 & 0.242 & \cellcolor{pink}0.194 & 0.053 & -0.142\\ 
\color{black}{$\sigma^{5,4}\pi_+^{2,2}\pi_-^{2,1}$} & \cellcolor{pink}4.399 & 4.641 & 0.242 & \cellcolor{pink}0.194 & 0.053 & -0.142\\ 
\color{black}{$\sigma^{5,4}\pi_+^{2,1}\pi_-^{2,1}\delta_-^{1,0}$} & 29.830 & 29.551 & -0.279 & 0.259 & 0.072 & -0.187\\ 
\color{black}{$\sigma^{5,4}\pi_+^{1,1}\pi_-^{2,2}\delta_-^{1,0}$} & 88.120 & 31.979 & -56.141 & 52.820 & 0.101 & -52.718\\ 
\color{black}{$\sigma^{5,3}\pi_+^{2,1}\pi_-^{3,1}\delta_-^{1,0}$} & 57.719 & 49.622 & -8.096 & 6.280 & 0.213 & -6.067\\ 
\color{black}{$\sigma^{5,4}\pi_+^{1,1}\pi_-^{3,1}\delta_-^{1,0}$} & 90.870 & 46.090 & -44.780 & 43.764 & 0.264 & -43.501\\ 
\color{black}{$\sigma^{5,3}\pi_+^{2,1}\pi_-^{2,1}\delta_-^{1,0}\phi_-^{1,0}$} & 286.259 & 250.639 & -35.620 & 22.000 & 7.920 & -14.081\\ 
\color{black}{$\sigma^{5,3}\pi_+^{1,1}\pi_-^{3,1}\delta_-^{1,0}\phi_-^{1,0}$} & 358.692 & 264.197 & -94.495 & 77.808 & 6.571 & -71.237\\ 
\hline
\end{tabular}
\caption{MAEDs between GTO and FEM energies in m$E_h$ for electronic configurations (following our shorthand notation) of the S atom in the fully uncontracted basis sets with real and complex orbitals. Configurations highlighted in magenta are equal in the real and complex basis and entries in light red belong to configurations that feature negative BSTEs.}
\label{tab:S-mean-differ}
\end{table*}

\begin{table*}
\centering
\small
\begin{tabular}{c|S[table-format=4.3]S[table-format=4.3]S[table-format=4.3]|S[table-format=4.3]S[table-format=4.3]S[table-format=4.3]}
\hline
 & \multicolumn{3}{|c}{aug-cc-pVTZ} & \multicolumn{3}{|c}{AHGBSP3-9}\\
\hline \hline
state & {Real} & {Complex} & {Change} & {Real} & {Complex} & {Change}\\ 
\hline \hline
\color{magenta}{$\sigma^{5,4}\pi_+^{2,2}\pi_-^{2,2}$} & 5.634 & 5.634 & 0.000 & 0.033 & 0.033 & 0.000\\ 
\color{black}{$\sigma^{5,5}\pi_+^{2,2}\pi_-^{2,1}$} & \cellcolor{pink}5.057 & 5.306 & 0.250 & \cellcolor{pink}0.223 & 0.031 & -0.193\\ 
\color{black}{$\sigma^{5,5}\pi_+^{2,1}\pi_-^{2,2}$} & \cellcolor{pink}5.057 & 5.306 & 0.250 & \cellcolor{pink}0.223 & 0.031 & -0.193\\ 
\color{black}{$\sigma^{5,4}\pi_+^{2,1}\pi_-^{2,2}\delta_-^{1,0}$} & \cellcolor{pink}44.993 & 44.948 & -0.045 & \cellcolor{pink}0.095 & 0.076 & -0.019\\ 
\color{black}{$\sigma^{5,5}\pi_+^{2,1}\pi_-^{2,1}\delta_-^{1,0}$} & 40.283 & 40.028 & -0.255 & 0.219 & 0.046 & -0.172\\ 
\color{black}{$\sigma^{5,4}\pi_+^{2,1}\pi_-^{3,1}\delta_-^{1,0}$} & 62.755 & 56.577 & -6.178 & 6.355 & 0.200 & -6.155\\ 
\color{black}{$\sigma^{5,4}\pi_+^{2,1}\pi_-^{2,1}\delta_-^{1,0}\phi_-^{1,0}$} & 504.280 & 462.441 & -41.840 & 20.978 & 8.477 & -12.500\\ 
\color{black}{$\sigma^{5,3}\pi_+^{2,1}\pi_-^{3,1}\delta_-^{1,0}\phi_-^{1,0}$} & 529.770 & 478.289 & -51.481 & 27.293 & 7.764 & -19.530\\ 
\color{black}{$\sigma^{6,3}\pi_+^{2,1}\pi_-^{3,1}\delta_-^{1,0}$} & 103.840 & 96.533 & -7.307 & 8.025 & 0.722 & -7.303\\ 
\hline
\end{tabular}
\caption{MAEDs between GTO and FEM energies in m$E_h$ for electronic configurations (following our shorthand notation) of the Cl atom in the fully uncontracted basis sets with real and complex orbitals. Configurations highlighted in magenta are equal in the real and complex basis and entries in light red belong to configurations that feature negative BSTEs.}
\label{tab:Cl-mean-differ}
\end{table*}

\begin{table*}
\centering
\small
\begin{tabular}{c|S[table-format=4.3]S[table-format=4.3]S[table-format=4.3]|S[table-format=4.3]S[table-format=4.3]S[table-format=4.3]}
\hline
 & \multicolumn{3}{|c}{aug-cc-pVTZ} & \multicolumn{3}{|c}{AHGBSP3-9}\\
\hline \hline
state & {Real} & {Complex} & {Change} & {Real} & {Complex} & {Change}\\ 
\hline \hline
\color{magenta}{$\sigma^{5,5}\pi_+^{2,2}\pi_-^{2,2}$} & 5.710 & 5.710 & 0.000 & 0.016 & 0.016 & 0.000\\ 
\color{magenta}{$\sigma^{6,4}\pi_+^{2,2}\pi_-^{2,2}$} & 46.060 & 46.060 & 0.000 & 1.752 & 1.752 & 0.000\\ 
\color{black}{$\sigma^{6,5}\pi_+^{2,2}\pi_-^{2,1}$} & \cellcolor{pink}45.779 & 46.033 & 0.254 & \cellcolor{pink}2.119 & 2.173 & 0.054\\ 
\color{black}{$\sigma^{6,5}\pi_+^{2,1}\pi_-^{2,2}$} & \cellcolor{pink}45.779 & 46.033 & 0.254 & \cellcolor{pink}2.119 & 2.173 & 0.054\\ 
\color{black}{$\sigma^{5,5}\pi_+^{2,1}\pi_-^{2,2}\delta_-^{1,0}$} & \cellcolor{pink}59.510 & 59.489 & -0.021 & \cellcolor{pink}0.062 & 0.046 & -0.016\\ 
\color{black}{$\sigma^{5,4}\pi_+^{2,1}\pi_-^{3,2}\delta_-^{1,0}$} & 95.142 & 75.856 & -19.286 & 20.740 & 0.192 & -20.548\\ 
\color{black}{$\sigma^{5,4}\pi_+^{2,1}\pi_-^{2,2}\delta_-^{1,0}\phi_-^{1,0}$} & 765.591 & 719.422 & -46.169 & 20.442 & 9.677 & -10.765\\ 
\color{black}{$\sigma^{5,4}\pi_+^{2,1}\pi_-^{3,1}\delta_-^{1,0}\phi_-^{1,0}$} & 779.603 & 717.280 & -62.323 & 26.642 & 8.358 & -18.284\\ 
\color{black}{$\sigma^{6,4}\pi_+^{2,1}\pi_-^{3,1}\delta_-^{1,0}$} & 129.817 & 122.196 & -7.621 & 8.400 & 0.764 & -7.635\\ 
\hline
\end{tabular}
\caption{MAEDs between GTO and FEM energies in m$E_h$ for electronic configurations (following our shorthand notation) of the Ar atom in the fully uncontracted basis sets with real and complex orbitals. Configurations highlighted in magenta are equal in the real and complex basis and entries in light red belong to configurations that feature negative BSTEs.}
\label{tab:Ar-mean-differ}
\end{table*}

\section{Summary and Conclusions}
\label{sec:summary}

We have presented an implementation of complex Gaussian-type orbital (GTO) basis functions for atoms and linear molecules in a parallel magnetic field.
In these cases the integrals turn out to be real, even though the basis is complex, and the implementation only requires additional unitary transformations before and after the Fock build, sitting on top of the normal real-basis machinery.
We applied this method to revisiting our earlier study of \citep{Aastroem2023_JPCA_10872}, and studying the low-lying configurations of the H--Ar atoms in magnetic fields up to $0.6B_0$.
We saw that the complex orbitals eliminated the problems encountered in our earlier study, so that the remaining errors are purely attributable to basis set truncation.
This development unlocks the cost-efficient optimization of systematic basis sets for finite magnetic fields, which is the topic of upcoming work from our group.

\section*{Author contributions}
H.\AA{}.: Software, Investigation, Formal analysis, Visualization, Writing -- original draft, Funding acquisition.
S.L.: Conceptualization, Methodology, Software, Supervision, Writing -- review \& editing, Funding acquisition.

\section*{Conflicts of interest}
There are no conflicts to declare.

\section*{Data availability}
The data supporting this article are included in the article and its electronic supplementary information.

\section*{Acknowledgements}
H.\AA{} thanks the Finnish Society for Sciences and Letters for financial support.
S.L. thanks the Academy of Finland for financial support under project numbers 350282 and 353749.



\balance

\bibliography{citations,mfields}
\bibliographystyle{rsc/rsc}

\end{document}